\documentclass[jamsb]{jaums}
\usepackage{setspace}
\theoremstyle{thmit} 

\theoremstyle{thmrm} 

\newtheorem*{oldproof}{Proof}

\usepackage{latexsym}
\usepackage{amsmath}
\usepackage{amsbsy}
\usepackage{amssymb}
\usepackage{epsfig}
\usepackage{graphicx}
\usepackage{graphicx}
\newtheorem{Shells}{Shell-Crossing theorem}[section]

\begin{document}


\title{Initial value formalism for Lemaitre-Tolman-Bondi collapse}

\author{P. D. Lasky}
\address{Centre for Stellar and Planetary Astrophysics\\
		School of Mathematical Sciences, Monash University\\
		Wellington Rd, Melbourne 3800, Australia}
\author{A. W. C. Lun}
\addressmark{1}
\author{R. B. Burston}
\addressmark{1}\address{Max Planck Institute for Solar System Research, 37191 Katlenburg-Lindau, Germany}

\date{January 10, 2007}
\maketitle

	\begin{abstract}
				Formulating a dust filled spherically symmetric metric utilizing the $3+1$ formalism for general relativity, we show that the metric coefficients are completely determined by the matter distribution throughout the spacetime.  Furthermore, the metric describes both inhomogeneous dust regions and also vacuum regions in a single coordinate patch, thus alleviating the need for complicated matching schemes at the interfaces.  In this way, the system is established as an initial-boundary value problem, which has many benefits for its numerical evolution. We show the dust part of the metric is equivalent to the class of Lemaitre-Tolman-Bondi (LTB) metrics under a coordinate transformation.  In this coordinate system, shell crossing singularities (SCS) are exhibited as fluid shock waves, and we therefore discuss possibilities for the dynamical extension of shell crossings through the initial point of formation by borrowing methods from classical fluid dynamics.   This paper fills a void in the present literature associated with these collapse models by fully developing the formalism in great detail.  Furthermore, the applications provide examples of the benefits of the present model.  
	\end{abstract}

		
\section{Introduction}

It is well established that naked singularities arise from the gravitational collapse of inhomogeneous, spherically symmetric dust spheres in general relativity \cite{yodzis73,hagen74,christodoulou84,newman86}.  One type of these are shell-crossing singularities (SCS), which are not perceived to violate cosmic censorship due to their weak nature (see e.g. \cite{szekeres99} and references therein).  However, difficulties often arise in the interpretation of the physical nature of SCS due to the four-dimensional construction of spacetime solutions usually having little or no resemblance to our intuitive perception of the universe.  Therefore, we utilize the $3+1$ formalism for general relativity to model the spacetime as an initial value problem in a single coordinate patch.  In this way the SCS are akin to shock waves in fluid mechanics, and therefore our perception and understanding of the solutions are greatly enhanced.    

The $3+1$ formalism involves specifying data on an initial Cauchy hypersurface, and subsequently evolving through time.  The ten Einstein field equations (EFEs) are decomposed into four constraint equations which must be satisfied on every spacelike hypersurface, and six evolution equations.  Furthermore, they can be supplemented by the integrability conditions for the EFEs, which are the Bianchi identities.  The once contracted Bianchi identities in four dimensional spacetimes provide 16 conditions, whilst the twice contracted provide the remaining four pieces of information, which are the conservation of energy-momentum equations.  

We apply the $3+1$ formalism for the EFEs and Bianchi identities to the case of a spherically symmetric sphere of inhomogeneous dust.  This article acts to fill a void in the literature created by \cite{lasky06b,lasky07} by deriving the formalism in full detail.  Furthermore, while \cite{lasky06b} and \cite{lasky07} used more general fluids, analytic solutions for the equations derived therein are extremely difficult to find, and hence the equations are awkward to interpret.  It is therefore pertinent to simplify the fluid in order to extract analytic solutions such that the full strength of the formalism can be displayed.  The dust application we present in this article is solved analytically, and hence a greater understanding of the structure of the equations is gained.  One of our main results from the dust section is that the metric coefficients are completely determined by the energy-momentum fields.  Furthermore, to determine the matter, and hence the entire spacetime, only the specification of the matter distribution on the initial hypersurface is required.  

We show our solution under a coordinate transformation is equivalent to the class of Lemaitre-Tolman-Bondi (LTB) solutions \cite{lemaitre33,tolman34,bondi47}.  Furthermore, the line element is such that an exterior vacuum region, namely the Schwarzschild spacetime expressed in a generalized form of the Painleve-Gullstrand (PG) coordinates \cite{painleve21,gullstrand22}, is described using the same coordinate patch.  In this way, the work is related to that of \cite{adler05} and \cite{gautreau95} who, in a single coordinate patch describe the Oppenheimer-Snyder (OS) collapse \cite{oppenheimer39}, i.e. a Friedmann-Robertson-Walker (FRW) interior joined to an exterior Schwarzschild spacetime.  As the FRW spacetime belongs to the class of LTB spacetimes, our work is a generalization of \cite{adler05} and \cite{gautreau95}.  

The evolution of the system is determined by two, first order equations in the matter fields.  The nature of these equations is such that multi-valued solutions in the mass can develop given smooth initial data.  These are equivalent to the SCS discussed, and we analyse the initial conditions necessary for these to occur.  Furthermore, we discuss a method for extending these SCS beyond their point of initial formation utilizing classical fluid dynamics methods.  In this way, we show when a globally naked SCS forms (i.e. beyond the apparent horizon) it will evolve to become a locally naked singularity, and eventually evolve into a Schwarzschild singularity at $r=0$.     

As all aspects of the solutions we present can be solved explicitly in terms of the initial data, the authors believe this formulation is of great value for testing numerical evolution codes.  In particular, the evolution of such a system contains many complications, including the formation of SCSs, as well as the interface between the two regions of the spacetime.  Numerical codes are extremely well developed to handle shock waves from classical fluid dynamics, and the same methods can be employed in the relativistic regime.  However, the evolution of the interface between the two regions may be of more primary concern.  In particular, the advantage of having both regions in a single coordinate patch implies no extra work in the code is required at the moving boundary.  One simply sets up the initial Cauchy data, with the matter terms going to zero at some finite radius, and the evolution will naturally take care of the free boundary.

The structure of the paper is as follows.  In section \ref{Derive} we derive the general form of the LTB metric and analyse the solution.  In section \ref{IC} we look at the initial and boundary conditions required for the specification of the spacetime, and also look at the formation of the apparent horizon.  In \ref{MBS} we reduce the system to the marginally bound case and derive the initial conditions that provide shock waves.  Finally, in section \ref{shock} we look at a specific example of the evolution of the system with a shock wave. 

Geometrized units are employed throughout whereby $G=c=1$.  Greek indices run from $0\ldots3$ and Latin from $1\ldots3$.  The Einstein summation convention is used, index conventions follow \cite{misner73} and $d\Omega^{2}$ is the line element for the two-sphere.

\section{The Metric}		\label{Derive}
Here, we pedagogically derive the spacetime metric to be used for the remainder of the article.  This metric is beneficial as it describes both an interior dust region as well as an exterior vacuum region without the need for complicated matching schemes across the interface.  In particular, both regions are represented not by two separate line elements patched together, but as a single, all encompassing, line element.  In this way, the metric we derive is a generalization of the PG vacuum solution to include dust in {\it any} region of the spacetime desired.   
   
The derivation of the metric utilizes the $3+1$ formalism for general relativity.  This is a powerful method that allows the resulting spacetime to be represented as an initial value problem.  In this way, initial Cauchy data is prescribed and the subsequent evolution depends continuously on the initial data.  Although analytic solutions are provided here, the authors believe the resulting class of spacetimes provides excellent test-beds for numerical schemes; in particular for gravitational collapse.

\subsection{ADM equations}\label{ADM}
The decomposition of the EFEs into a $3+1$ system is otherwise known as the ADM formalism \cite{arnowitt62}.  Solving the equations in this form establishes a metric appropriate to an initial value problem.  

Consider a four-dimensional spacetime foliated with three-dimensional spacelike hypersurfaces.  We denote the four-coordinates with $x^{\mu}:=\left(t,x^{i}\right)$, and a spherically symmetric line element can be reduced for dust, without loss of generality, to 
\begin{align}
ds^2=-\left(\alpha^2-\mathcal{U}^{2}\beta^{2}\right)dt^2+2\mathcal{U}^2\beta dtdr+\mathcal{U}^{2}dr^2+r^2d\Omega^2.	\label{genmetric2}
\end{align}
where $\alpha=\alpha(t,r)$ is the lapse function, $\mathcal{U}=\mathcal{U}(t,r)$ and $\beta=\beta(t,r)$ is the radial component of the shift vector.  

The energy-momentum tensor for dust is given by
\begin{align}
T_{\mu\nu}=\rho n_{\mu}n_{\nu},
\end{align}
where $\rho$ is the energy-density, and $n^{\mu}$ is the normal vector field tangent to the fluid lines.  We demand three properties for the normal vector, given by two equations
\begin{align}
n^{\alpha}n_{\alpha}=-1\qquad\textrm{and}\qquad n_{[\mu}\nabla_{\nu}n_{\sigma]}=0,\label{ns}
\end{align}
where $\nabla_{\mu}$ is the unique four-covariant derivative operator.  The first equation in (\ref{ns}) implies the vector $n^{\mu}$ is both timelike and normalized, and the second equation implies the normal vector is hypersurface forming.   Utilizing the line element (\ref{genmetric2}) and both equations in (\ref{ns}), one can show the normal vector in component form is given by
\begin{align}
n^{\mu}=\frac{1}{\alpha}\left(1,-\beta,0,0\right).\label{normal}
\end{align}     
Using the normal vector, the EFEs, $G_{\mu\nu}=8\pi T_{\mu\nu}$, can be decomposed into four constraint equations which must be satisfied on each spacelike hypersurface.  These are the Hamiltonian constraint
\begin{align}
{^3R}+\frac{2}{3}K^2-A_{ij}A^{ij}=16\pi\rho,			\label{Hamiltonian}
\end{align}
where ${^{3}R}$ is the three-Riemann scalar, $A_{ij}$ and $K$ are the trace-free part and trace of the extrinsic curvature respectively.  There are three momentum constraints
\begin{align}
D^jA_{ij}=\frac{2}{3}D_iK,					\label{Momentum}
\end{align} 
where $D_{i}$ is the unique three-covariant derivative operator.  The remaining six equations are the evolution equations,
\begin{align}
2\mathcal{L}_{n}K=&K^{2}+\frac{3}{2}A_{ij}A^{ij}+\frac{1}{2}{^{3}R}-\frac{2}{\alpha}D^{i}D_{i}\alpha,                                      \label{Ev1}\\
\mathcal{L}_{n}A_{ij}=&\frac{1}{3}KA_{ij}-2A_{ik}{A_{j}}^{k}+{^{3}R}_{ij}-\frac{1}{3}\perp_{ij}{^{3}R}\nonumber\\
&\qquad\qquad\qquad-\frac{1}{\alpha}D_{i}D_{j}\alpha+\frac{1}{3\alpha}\perp_{ij}D^{k}D_{k}\alpha,                                     \label{Ev2}
\end{align}
where $\mathcal{L}_{n}$ is the Lie derivative operator with respect to the normal vector.  

Equations (\ref{Hamiltonian}-\ref{Ev2}) are the ADM equations, and these need to be supplemented with the conservation of energy-momentum equations
\begin{align}
\nabla_{\alpha}{T^{\alpha}}_{\mu}=0.
\end{align}
These can be decomposed giving the continuity equation which, for dust reduces to
\begin{align}
K=\mathcal{L}_n\left(\ln\rho\right),						\label{Euler}
\end{align}		
and the Euler equations which, in dust, are
\begin{align}
\rho\,\, D_{i}\left(\ln\alpha\right)=0.					\label{continuity}
\end{align}
As we are considering systems with non-vanishing energy density, equation (\ref{continuity}) along with the form of the normal vector (\ref{normal}), is enough to show the lapse function is only a function of the time coordinate.  Utilizing coordinate freedom, we set the lapse function to unity without loss of generality,
\begin{align} 
\alpha=1.
\end{align}
This physically implies that the dust particles are moving along timelike geodesics of the spacetime, which is a result of there being zero pressure.

\subsection{Governing equations}

We begin by putting the metric (\ref{genmetric2}), into the ADM system of equations.  Noting that the Lie derivative acting on a scalar function, $\psi$, is simply
\begin{align}
\mathcal{L}_n\psi&=n^{\alpha}\partial_{\alpha}\psi\nonumber\\
&=\frac{\partial\psi}{\partial t}-\beta\frac{\partial\psi}{\partial r},	\label{Liedef}
\end{align}
helps to recognize patterns in the systems of equations.  In particular, we immediately find the radial component of equation (\ref{Momentum}) reduces to
\begin{align}
\mathcal{L}_n\mathcal{U}=0.				\label{LieU}
\end{align}
This term, and its derivatives appear readily throughout the system of equations.  Substituting (\ref{LieU}) into the remaining equations implies the rest of the system reduces to a complicated set of differential equations in $\mathcal{U}$, $\rho$ and $\beta$.  

Applying various combinations of the $3+1$ EFEs results in being able to write the system algebraically in $\mathcal{U}$ and $\rho$, leaving only derivatives of the shift function $\beta$.  In particular, combining (\ref{Hamiltonian}) and (\ref{Ev2}) we find the derivatives of $\mathcal{U}$ cancel and a relation between $\rho$ and derivatives of the shift results in
\begin{align}						
r^2\mathcal{L}_n\beta=4\pi\int_{\sigma=0}^r\rho\left(t,\sigma\right)\sigma^2d\sigma:=M(t,r),					\label{rhocond}
\end{align}
where we have defined a ``mass'' function according to a Newtonian definition.  We note here that this function does not physically represent the mass of the system as a Newtonian volume element has been used.  However, this function arises in a natural way.  Furthermore, we will show that this is precisely the mass function commonly defined in the LTB spacetimes.  Furthermore, in a vacuum region of the spacetime where the energy density vanishes, the mass function simply becomes a constant.  We will see that this vacuum spacetime is given by the Schwarzschild spacetime in a generalization of the Painleve-Gullstrand coordinates (see section \ref{ext}).    
 
Adding equation (\ref{Ev1}) to six times the radial component of equation (\ref{Ev2}) enables derivatives of $\mathcal{U}$ to again cancel, resulting, after some algebra, in
\begin{align}
\mathcal{U}^2&=\frac{1}{1+\beta^2-2r\mathcal{L}_n\beta}.							\label{Ucond}
\end{align}
We now find that the final metric function $\mathcal{U}$ is related to the shift, and therefore, implicitly related to the energy density.  Thus for dust, the spherically symmetric metric in the form given by (\ref{genmetric2}) is entirely dependant on the initial matter field.  

Now by substituting (\ref{Ucond}) into (\ref{LieU}) we find a second order evolution equation solely on the shift
\begin{align}
0&=\mathcal{L}_n\left(\beta^2-2r\mathcal{L}_n\beta\right).							\label{condition}				
\end{align}

Summarily, the line element takes the form
\begin{align}
ds^2=-dt^2+\frac{\left(\beta dt+dr\right)^2}{1+\beta^2-2r\mathcal{L}_n\beta}+r^2d\Omega^2,			\label{LTourcoord}
\end{align}
where $\beta$ is given by a solution of equation (\ref{condition}) and is further constrained by the requirement that the signature of the spacetime be Lorentzian, $\beta^2-2r\mathcal{L}_n\beta>-1$.  

Due to the method of deriving the metric, we note that it is conducive to an initial value formulation.  In particular, the evolution of the system is governed by the second order partial differential equation (\ref{condition}).  To solve this equation one must pose boundary and initial data, which is discussed in section \ref{IC}.  


The physical realization this method has brought is due to the manner in which it was derived.  Many methods for solving the EFEs require simply solving for the full four-dimensional spacetime with minimal consideration of the physics until the solution has been found.  It is then difficult to decipher the physics due to the seemingly arbitrary nature of the coordinates.  By specifying spacelike hypersurfaces and allowing for the subsequent evolution of the system, we have reduced the problem to one that relates closer to the physical intuition grasped from everyday life.

\subsection{Gravitoelectromagnetism}
Another important set of equations which can be solved alongside the ADM equations (\ref{Hamiltonian}-\ref{Ev2}) and the conservation equations (\ref{Euler}-\ref{continuity}) are the remaining Bianchi identities.  In four dimensions one can go to the once contracted Bianchi identities without loss of generality.  By introducing the Weyl conformal curvature tensor, the once contracted Bianchi identities can be expressed in terms of the energy momentum tensor
\begin{align}
\nabla^\alpha C_{\alpha\mu\nu\sigma}=8\pi\left(\nabla_{[\nu}T_{\sigma]\mu}+\frac{1}{3}g_{\mu[\nu}\nabla_{\sigma]}{T_\alpha}^\alpha\right). \label{Bianchi}
\end{align}
The Weyl tensor can be decomposed into two spatial, trace free, gravito-electromagnetic (GEM) tensors, known as the electric and magnetic conformal curvatures, $E_{ij}$ and $B_{ij}$ respectively, where
\begin{align}
E_{ij}:&=C_{i\alpha j\beta}n^{\alpha}n^{\beta},\\ 
B_{ij}:&=\frac{1}{2}\varepsilon_{\alpha i\beta\gamma}{C^{\beta\gamma}}_{\delta j}n^\alpha n^\delta.
\end{align} 

The electric conformal curvature is a measure of the tidal forces present in the spacetime, analogous to the tidal tensor in Newtonian gravitation.  The magnetic conformal curvature has no Newtonian analogue, and is in some sense a measure of the intrinsic rotation associated with the spacetime \cite{ellis71}.  Therefore, while $B_{ij}=0$ in spherical symmetry, we note that it will have non-trivial contributions for spacetimes with less symmetries.  For example, one aim of the current program of research is to generalize this method to include the family of Robinson-Trautman solutions which contain gravitational radiation.  These spacetimes will begin with a non-zero magnetic curvature.  However one  in these scenarios that this contribution will be ``radiated away'' in the form of gravitational radiation such that the steady-state solution is spherically symmetric.  Furthermore, a radiating, axially-symmetric spacetime will have two components of the magnetic curvature.  One component will be associated with the intrinsic angular momentum of the spacetime, and will remain once the system has radiated to a steady-state.  The other component associated with the gravitational radiation will be radiated away in the vein associated with the Robinson-Trautman solutions.    

As mentioned, for spherically symmetric dust the magnetic curvature tensor vanishes and the only contribution to the Weyl tensor is from the electric curvature.  The trace-free nature of the electric curvature, as well as the spherical symmetry imposed, implies it has the simple form
\begin{align}
{E_i}^j=\textrm{diag}(-2\lambda,\lambda,\lambda),			\label{Eform}
\end{align}  
where $\lambda=\lambda(t,r)$ is the tidal forces associated with the spacetime.  Decomposing the Bianchi identities, given $B_{ij}=0$, yields two non-trivial equations
\begin{align}
&D^kE_{ki}=\frac{8\pi}{3}D_i\rho,\label{DE}\\
\mathcal{L}_nE_{ij}-\perp_{ij}E_{kl}A&^{kl}+5{A_{(i}}^kE_{j)k}-\frac{1}{3}E_{ij}K=4\pi\rho A_{ij}.\label{CurlB}
\end{align}
Equation (\ref{DE}) can be integrated by using the form of the electric curvature (\ref{Eform}), and by substituting the definition for the mass (\ref{rhocond}), to give the tidal forces for the spacetime
\begin{align}
\lambda=-\frac{r}{3}\frac{\partial}{\partial r}\left(\frac{M}{r^{3}}\right).\label{lambda}
\end{align}
We note that in a vacuum region, the mass function is constant, and the tidal forces therefore become
\begin{align}
\lambda_s=\frac{M_s}{r^3},						\label{schwla}
\end{align}
which is the familiar form known for the Schwarzschild spacetime.  These quantities will become important for the discussion of shock formation in section \ref{classical}.

The other non-trivial GEM equation (\ref{CurlB}) can be calculated using equations (\ref{rhocond}) and (\ref{Eform}), and also substituting (\ref{lambda}), and it simply gives an integrability check for the evolution equation which will be given for the mass function (see section \ref{crux}, in particular equation (\ref{dmdt})).



\subsection{Equivalence with LTB}\label{crux}
We have derived the reduced field equations (\ref{condition}), for the line element (\ref{LTourcoord}) that represents spherically symmetric dust.  We now show that this can be transformed into LTB coordinates $(\tau,R, \theta,\phi)$.

Let $r$ be a function of both the new time and new radial coordinates, i.e., $r=r(\tau,R)$, and also let $t=\tau$, then let
\begin{align}
\frac{\partial r(\tau,R)}{\partial \tau}=-\beta.			\label{newcoords}
\end{align}
In this new coordinate system, the normal vector has the form $n^{\mu^\prime}=(1,0,0,0)$, and the observer is now comoving with the fluid.  In these coordinates the Lie derivative of a scalar function is simply the derivative with respect to the new temporal coordinate, $\tau$, and the condition given by (\ref{condition}) becomes
\begin{align}
\frac{\partial}{\partial \tau}\left[\left(\frac{\partial r(\tau,R)}{\partial \tau}\right)^2+2r(\tau,R)\frac{\partial^2r(\tau,R)}{\partial \tau^2}\right]=0. 		\label{muhmuhmuh}	
\end{align}
This is a third order differential equation in the function $r(\tau,R)$, and can be integrated twice to yield
\begin{align}
\left(\frac{\partial r(\tau,R)}{\partial\tau}\right)^2=E(R)+\frac{2m(R)}{r(\tau,R)},					\label{LTusualcond}
\end{align}
where $E(R)$ and $m(R)$, known as the energy and Misner-Sharp mass \cite{misner64} respectively, are functions of integration.  Furthermore, rearranging the metric we find
\begin{align}
ds^2=-d\tau^2+\frac{\left(\frac{\partial r(\tau,R)}{\partial R}\right)^2}{1+E(R)}dR^2+r(\tau,R)^2d\Omega^2.		\label{LTusualcoord}
\end{align}
Equations (\ref{LTusualcoord}) and (\ref{LTusualcond}) are the standard form of the LTB metric \cite{lemaitre33,tolman34,bondi47}. 

The function $E(R)$, arbitrary up to the constraint $E(R)>-1$, is a measure of the energy of a shell at a radius $R$ \cite{szekeres99}.  Therefore, this function being equal to zero is equivalent to saying the particles at infinity have zero kinetic energy. 

It was found during the above transformation that this energy function can be expressed in terms of the shift function in our coordinates
\begin{align}    
E(R)=\beta^2-2r\mathcal{L}_n\beta,			\label{Efunction}
\end{align}
which is simply the first integral of equation (\ref{condition}).

Transferring back into the original coordinates, we note that the energy function is a function of $t$ and $r$.  Furthermore, substituting equation (\ref{Efunction}) back through (\ref{condition}), we see  
\begin{align}
\mathcal{L}_{n}E=0.
\end{align}
Now, substituting the mass defined in equation (\ref{rhocond}) back into (\ref{Efunction}) gives 
\begin{align}
\beta^{2}=E+\frac{2M}{r},\label{be}
\end{align}
from which we can further show this implies
\begin{align}
\mathcal{L}_{n}M=0.
\end{align}
Finally, the above equations can be put back through the line element, and we find the resulting system can be summarized by the metric and two constraint equations
\begin{subequations}\label{System}
\begin{align}
ds^{2}=\-dt^{2}+&\frac{\left(\sqrt{E+2M/r}\,\,dt+dr\right)^{2}}{1+E}+r^{2}d\Omega^{2}.             \label{mainmetric}\\
&\frac{\partial M}{\partial t}-\sqrt{E+\frac{2M}{r}}\frac{\partial M}{\partial r}=0                             \label{dmdt}\\
&\frac{\partial E}{\partial t}-\sqrt{E+\frac{2M}{r}}\frac{\partial E}{\partial r}=0                              \label{dedt}
\end{align}
\end{subequations}
Equations (\ref{dmdt}) and (\ref{dedt}) provide a coupled system of differential equations which are solved concurrently to determine the dynamics of the system.  

We note that the positive root of equation (\ref{be}) has been selected as this represents a collapsing model.  Choosing the negative root is equivalent to reversing the time coordinate, and therefore gives an expanding, cosmological type model.

\subsection{Vacuum Regions}\label{ext}
While we have derived this solution as a dust problem, we note that these coordinates also describe vacuum regions.  This is shown by simply letting the energy density vanish at some radii on the initial hypersurface.  This implies that the mass function becomes constant, and thus equation (\ref{dmdt}) is trivially satisfied.  Therefore, vacuum regions of the system are described by the line element (\ref{mainmetric}) and equation (\ref{dedt}) with $M=M_{s}\,\,\in\,\,{\bf R}$.  We describe this system as the ``generalized Painleve-Gullstrand'' (GPG) coordinates\footnote{A more detailed analysis of these vacuum coordinates will be presented in a future article.} as they reduce to the usual PG coordinates \cite{painleve21,gullstrand22} for the particular case of $E=0$.  Evaluating the Einstein tensor for the GPG metric we find it vanishes as expected.  This is therefore a vacuum solution of the field equations, and spherical symmetry along with Birkhoff's theorem implies this solution must be diffeomorphic to the Schwarzschild metric.  

However, the vacuum solution described by (\ref{mainmetric}) and (\ref{dedt}) is actually a family of solutions parametrized by the energy function.  Therefore, for all solutions of equation (\ref{dedt}), there will exist a coordinate transformation to the Schwarzschild metric, in coordinates $(T,r,\theta,\phi)$.  This transformation is given by solutions to the following coupled system of differential equations
\begin{subequations}					\label{Transform}
\begin{align}
\left(\frac{\partial t}{\partial T}\right)^{2}=&1+E,\label{Transforma}\\
\left(1-\frac{2M_{s}}{r}\right)\frac{\partial t}{\partial r}=&\sqrt{\frac{2M_{s}}{r}+E}.\label{Transformb}
\end{align}
\end{subequations}

To show this is a valid coordinate transformation amounts to showing that there exists a solution of (\ref{Transform}), which is done by checking the integrability conditions.  That is, by differentiating equation (\ref{Transforma}) with respect to the radial coordinate, and equation (\ref{Transformb}) with respect to the temporal coordinate, one can show the partial derivatives commute providing equation (\ref{dedt}) is satisfied.  Therefore, providing the energy function is a solution of (\ref{dedt}), the coordinate transformation is valid, and applying the coordinate transformation yields
\begin{align}
ds^{2}=-\left(1-\frac{2M_{s}}{r}\right)dT^{2}+\frac{dr^{2}}{1-2M_{s}/r}+r^{2}d\Omega^{2},
\end{align}
which is exactly the Schwarzschild metric.

\section{Initial and Boundary Conditions}\label{IC}
We can now establish the initial and boundary data required to solve the system of equations that describe the spacetime.  We are required to specify an energy density for the initial hypersurface.  Due to the formulation of the solution, this energy density can take any form, including vanishing for finite regions.  For example, if one wanted to study the gravitational collapse of a spherical body, then one specifies an appropriate energy density out to a finite radius, at which point the energy density is allowed to vanish beyond that radii.  However, one can also study the collapse of concentric shells of matter; for example by specifying successive Heaviside step functions for the initial density.  The robustness of the formalism allows for the study of any initial configuration of spherically symmetric dust.  

Whatever form the initial energy density takes, we can evaluate the initial mass function through the definition given by equation (\ref{rhocond}).  For vacuum regions, this mass function reduces to a constant.  We therefore have the initial conditions for the mass function throughout the initial spacelike hypersurface.  

The data for the energy function is a little more tricky.   By expressing equation (\ref{dmdt}) in terms of the energy function, and substituting equation (\ref{rhocond}), after some algebra we find
\begin{align}
E(0,r\le r_{\partial})=\left(\frac{\int_0^r\left.\frac{\partial\rho}{\partial t}\right|_{t=0}\sigma^2d\sigma}{\rho(0,r) r^2}\right)^2-\frac{8\pi}{r}\int_0^r\rho(0,\sigma)\sigma^2d\sigma.
\end{align}
Therefore, the energy function is found by specifying two terms, namely
\begin{align}
\left.\frac{\partial\rho}{\partial t}\right|_{t=0}\qquad\textrm{and}\qquad\rho(0,r).
\end{align}
While this works for the non-vacuum regions, one still has the freedom to choose the form of the energy function for the vacuum regions.  However, if all regions of the spacetime are to be described by a single coordinate patch, then one can employ continuity of the energy function across the interface of the matter filled and vacuum regions to determine the boundary conditions for the vacuum region.


\section{Apparent horizon}

No discussion of gravitational collapse is complete without contemplating the apparent horizon.  We shall now show that the coordinate system used herein allows for the description of the apparent horizon in a clear and concise manner.  

The apparent horizon is defined as the boundary of the closure of the union of all trapped regions on a Cauchy surface \cite{wald84}.  An alternate, but equivalent definition is given by the surface with a vanishing expansion factor, $\Theta$, which is defined as the divergence of a congruence of null geodesics
\begin{align}
\Theta:=\nabla_{\alpha}k^{\alpha}.\label{Theta}
\end{align}  
Here $k^{\mu}$ is a null vector which is everywhere tangential to the congruence of radial null geodesics.  To derive this null vector, we must look at the equations of motion describing the null geodesics in the spacetime.  These are given by the Lagrangian, $\mathcal{L}$, and the Euler-Lagrange equations,
\begin{align}
2\mathcal{L}\left(\dot{x}^{\mu},x^{\mu}\right)&=\frac{1}{1+E}\left[-\left(1-\frac{2M}{r}\right)\dot{t}^{2}+2\sqrt{\frac{2M}{r}+E}\dot{t}\dot{r}+\dot{r}^{2}\right]=0,\label{Lagrangian}\\
&0=\frac{d}{d\lambda}\frac{\partial\mathcal{L}}{\partial\dot{t}}-\frac{\partial\mathcal{L}}{\partial t}\qquad\textrm{and}\qquad0=\frac{d}{d\lambda}\frac{\partial\mathcal{L}}{\partial\dot{r}}-\frac{\partial\mathcal{L}}{\partial r},\label{EulerLagrange}
\end{align}
where a dot denotes differentiation with respect to the affine parameter along the geodesics, $\lambda$.  Furthermore, we have used spherical symmetry to imply that the apparent horizon will only depend on the radial and temporal coordinates, i.e. $\dot{\theta}=\dot{\phi}=0$.  

The three equations (\ref{Lagrangian},\ref{EulerLagrange}) can now be integrated to solve for $\dot{t}$ and $\dot{r}$ in terms of a single constant of integration and the metric coefficients.  The null vector $k^{\mu}:=\dot{x}^{\mu}$ can then be expressed 
\begin{align}
k^{\mu}=\left[\sqrt{1+E}\,\,,1+E-\sqrt{1+E}\sqrt{\frac{2M}{r}+E}\,\,,0,0\right],\label{k}
\end{align}
where the constant of integration has been scaled to unity without loss of generality.   Taking the divergence of this quantity, as indicated by equation (\ref{Theta}), one finds the expansion factor vanishes along the surface given by the implicit parametric equation
\begin{align}
r\left(t\right)=2M\left(t,r\left(t\right)\right).
\end{align}
This remarkably simple form is true for all choices of initial conditions.  Furthermore, at the interface between the matter and vacuum regions, the mass function simply becomes the Schwarzschild mass, and the horizon reduces to the familiar event horizon in the Schwarzschild spacetime.

\section{Marginally Bound Solution}\label{MBS}
For the remainder of the article we only consider the case where the energy function, $E(R)$, vanishes.  Known as the marginally bound case\footnote{$E(R)<0$ is the bound case and $E(R)>0$ is the unbound case.}, this model provides a simpler example than the non-zero energy case.  The marginally bound system is described by the line element and single condition on the mass
\begin{subequations}
\begin{align}
ds^{2}=-dt^{2}&+\left(\sqrt{\frac{2M}{r}}dt+dr\right)^{2}+r^{2}d\Omega^{2} ,           \label{metric}\\
&\frac{\partial M}{\partial t}-\sqrt{\frac{2M}{r}}\frac{\partial M}{\partial r}=0   . \label{Masscond}
\end{align}
\end{subequations}

Equation (\ref{Masscond}) is a first order, quasi-linear partial differential equation which governs the dynamics of the system.  The fact that this can easily be written in conservation form is important to the numerical analysis of the system.  In particular, systems that can be written in conservation form are preferred since the convergence (if it exists) to weak solutions of the equations  are guaranteed by the Lax-Wendroff theorem \cite{lax60}. 

The general solution to (\ref{Masscond}) is given implicitly by
\begin{align}
M=\mathcal{F}\left(\frac{2}{3}r^{3/2}+t\sqrt{2M}\right),		\label{MassGenSoln}
\end{align}
where $\mathcal{F}$ is a function of integration.  To find the particular solution, and hence determine the dynamics of the system, one simply requires the input of an initial mass distribution.  This initial data can be expressed in terms of the energy density, which is translated into an initial mass via the definition (\ref{rhocond}).  

The nature of equation (\ref{MassGenSoln}) is such that given smooth initial data, the solution can evolve to be multi-valued for the mass function, violating physical intuition regarding the behaviour of mass.  These multi-valued solutions correspond to SCS in general relativity, and are also described by mathematical shock waves in this formalism.  The solution prior to the formation of these shocks is described by the classical solution of the differential equation (\ref{Masscond}) while after the formation of the shock the solution can still be evolved utilizing a weak solution.


\subsection{Classical Solution} \label{classical}
By analyzing the characteristics of (\ref{Masscond}), we can determine the point at which the classical solution fails, and also the initial conditions required for this to occur at some point within the collapse process.  

The characteristics of equation (\ref{Masscond}) can be represented by introducing a parameter $\xi$, such that
\begin{align}
t=t(\xi),\,\,r=r(\xi)\,\,\text{and}\,\, M=M(\xi).
\end{align}
Furthermore, by defining $t(0)=0$ without loss of generality, and defining a new parameter $s:=r(0)$, we see  $M\left(t(0),r(0)\right)=M\left(0,s\right):=M_{0}(s)$.  Now, the characteristics are given by the parametric equations 
\begin{subequations}\label{Charset}
\begin{align}
M=M_{0}(s),&\,\,\,\,\,\,\,\,\,\,\,\,\,\,\, t=\xi,\\
r^{3/2}=s^{3/2}-&\frac{3}{2}\xi\sqrt{2M_{0}(s)}.
\end{align}
\end{subequations}
The solution becoming multi-valued corresponds to the intersection of the characteristic curves, which is interpreted as when the transformation $(\xi,s)\rightarrow(t,r)$ becomes non-invertible.  That is, the determinant of the Jacobian of transformation vanishes.  Therefore, to find conditions for the formation of the shock, we must solve
\begin{align}
\mathcal{J}:=\left|\begin{array}{cc}
\frac{\partial t}{\partial\xi} & \frac{\partial t}{\partial s}\vspace{0.2cm}\\
\frac{\partial r}{\partial\xi} & \frac{\partial r}{\partial s}
\end{array}\right|=0.
\end{align}
It is trivial to show from equations (\ref{Charset}) that
\begin{align}
\mathcal{J}=0\,\,\, \Leftrightarrow\,\,\,\frac{\partial r}{\partial s}=0.
\end{align}
This condition is equivalent to 
\begin{align}
t=&\frac{\sqrt{2sM_{0}}}{M_{0}^{\prime}},			       		              \label{Cond1}\\
r=&s^{1/3}\left(1-\frac{3M_{0}}{M_{0}^{\prime}}\right)^{2/3},                      \label{Cond2}
\end{align}
where a prime denotes differentiation with respect to $s$.  We are only concerned with shocks that occur in the first quadrant of the $\left(t,r\right)$ plane.  Therefore, we must determine the conditions on $M_{0}$ and $M_{0}^{\prime}$ which imply the Jacobian vanishes.  

As $s=r(t=0)$, we see $s\in\left[\left.0,\infty\right)\right.$.  Furthermore, $\rho\ge0$ and $M_{0}(s)$ is defined by letting $t=0$ in equation (\ref{rhocond}), implying $M_{0}(s)\ge0$ for all $s$.  Therefore, by demanding $t>0$ in (\ref{Cond1}), we see
\begin{align}
M_{0}^{\prime}>0. 
\end{align}
This will only {\it not} hold true when $\rho(0,r)=0$ for some finite range of $r$, as this will imply $M_{0}^{\prime}=0$.  Therefore, the characteristics emanating from radii for which there is a vanishing energy density will never cross, as expected.  

Now, demanding $r>0$ in equation (\ref{Cond2}) implies, after some manipulation
\begin{align}
\frac{d}{ds}\left(\frac{M_{0}}{s^{3}}\right)>0.                      \label{ineq}
\end{align}
Thus, shocks occur in the first quadrant of the $\left(t,r\right)$ plane if for any finite range of $s\in[0,\infty)$, the inequality in (\ref{ineq}) is satisfied.    

Finally, letting $t=0$ in (\ref{lambda}) implies the ``tidal force'' at $t=0$ is
\begin{align}
\lambda(0,s)=-\frac{s}{3}\frac{d}{ds}\left(\frac{M_{0}}{s^{3}}\right),
\end{align}
and considering $s>0$, the inequality in (\ref{ineq}) is equivalent to
\begin{align}
\lambda(0,r)<0.
\end{align}
Summarily, we have proved the following theorem:
\begin{Shells}
SCS occur in marginally bound LTB collapse if and only if $\lambda(0,r)<0$ for some finite range of $r\in\left[\left.0,\infty\right)\right.$. 
\end{Shells}
Conversely, if $\lambda(0,r)\ge0$ for all $r$, then shock waves will not form.  Furthermore, the initial time for the shock to begin, denoted $t_{s}$, can be determined solely from the initial conditions by taking the minimum of (\ref{Cond1}),
\begin{align}
t_{s}=\sqrt{2}\min\left\{\frac{\sqrt{sM_{0}}}{M_{0}^{\prime}}\right\}.
\end{align}

\subsection{Weak Solution}
The solution is satisfied classically until the point where the characteristics first cross.  This is the initial point of the shock surface, and beyond this point the solution to equation (\ref{Masscond}) is only given by a weak solution.  The analysis of the weak solution is made simpler by putting the equation into conservation form.  This is done by rescaling the radial coordinate such that $\hat{r}:=r^{3/2}$, implying
\begin{align}
\frac{\partial M}{\partial t}+\frac{\partial}{\partial\hat{r}}f(M)=0,
\end{align}
where $f(M):=-\sqrt{2}\,\,M^{3/2}$.  The weak solution is defined as follows: consider a test function $\psi\in{\bf C}^{1}$ which vanishes everywhere outside a rectangle defined by $0\le t\le T$ and $a\le\hat{r}\le b$, and also on the lines $t=T$, $a=\hat{r}$ and $b=\hat{r}$ (for some $a$, $b$ and $T\in{\bf R}$).  A weak solution is defined to be a solution of 
\begin{align}
\int^{T}_{t=0}\int^{b}_{\hat{r}=a}\Big(M\frac{\partial\psi}{\partial t}+&f(M)\frac{\partial\psi}{\partial\hat{r}}\Big)d\hat{r}dt\nonumber\\
+&\int^{b}_{\hat{r}=a}M_{0}(\hat{r})\psi(0,\hat{r})d\hat{r}=0,
\end{align}
for all test functions $\psi$.

The weak solution contains a world-line whereby the solution contains a jump discontinuity which is the shock surface.  This world-line can be expressed parametrically in terms of the time coordinate, {\it i.e.} $\hat{r}_{s}(t)$.  Either side of the shock, the solution is satisfied by the usual classical solution discussed in section \ref{classical}.

A method for evolving shocks beyond the initial point of formation, utilized in many classical scenarios, is to introduce a viscosity term into the differential equation, 
\begin{align}
\frac{\partial M}{\partial t}+\frac{\partial}{\partial\hat{r}}f(M)=\varepsilon\frac{\partial^{2}M}{\partial\hat{r}^{2}}.
\end{align}
Here, $\varepsilon\rightarrow0^{+}$ near $\hat{r}=\hat{r}_{s}(t)$, and is zero elsewhere.  This has the effect of {\it smoothing out} the shock into a travelling wave solution.  Despite the spacetime being pressureless, we can imagine when particles begin to get extremely close to one another, as is the case just before a shock begins to occur, a force would begin to play a role that kept these particles from getting too close.    

Utilizing the viscosity term in the differential equation, we can derive a condition on the shock known as the Rankine-Hugionot condition, which gives the velocity of the shock (see for e.g. \cite{smoller83}).  This is given by
\begin{align}
\hat{V}_{s}:=\frac{d\hat{r}_{s}}{dt}=\frac{\left[f(M)\right]^{+}_{-}}{\left[M\right]^{+}_{-}},                     \label{RHwithrhat}
\end{align}
where 
\begin{align}
\left[\ldots\right]^{+}_{-}:=\lim_{r\rightarrow r_{s}^{+}}- \lim_{r\rightarrow r_{s}^{-}}.
\end{align} 
This result can be transformed back into the radial coordinate utilized in the metric, and the velocity of the shock is found to be
\begin{align}
V_{s}:=\frac{dr_{s}}{dt}=\frac{-2\sqrt{2}}{3\sqrt{r}}\frac{\left[M^{3/2}\right]^{+}_{-}}{\left[M\right]^{+}_{-}}.                      \label{RHwithr}
\end{align}
We note that the smoothing either side of the shock necessarily implies that the mass is still a monotonically increasing function in the positive radial direction.  Therefore, $V_{s}$ must necessarily be negative, implying the shock travels in the direction towards $r=0$.  Furthermore, once formed there is no mechanism to stop the evolution of the shock, and therefore during the evolution it will reach $r=0$, at which point the shock will cease to exist and the classical solution will be regained.  Thus, we see that even if a shock forms as a global naked singularity, it will evolve to the end state of gravitational collapse as a black hole.

\section{Shock Example}\label{shock}
We wish to highlight the formation and evolution of a shock wave by way of example.  In particular, specifying an initial matter distribution such that the initial tidal force, $\lambda(0,r)<0$ for a finite range of $r$, ensures a shock will develop throughout the evolution.  We define an initial energy-density distribution by the following piecewise continuous function
\begin{align}
\rho_{0}=\left\{\begin{array}{cc}
\frac{2}{3}\cos\frac{2\pi r}{3r_{\partial_{0}}}-\frac{2}{3}\cos\frac{2\pi r}{r_{\partial_{0}}}+1 & r\le r_{\partial_{0}}\\
0 & r>r_{\partial_{0}}
\end{array}\right.. 			\label{initdensity}
\end{align}
\begin{figure}
		\begin{center}
		\includegraphics[height=0.45\textwidth,width=0.45\textwidth]{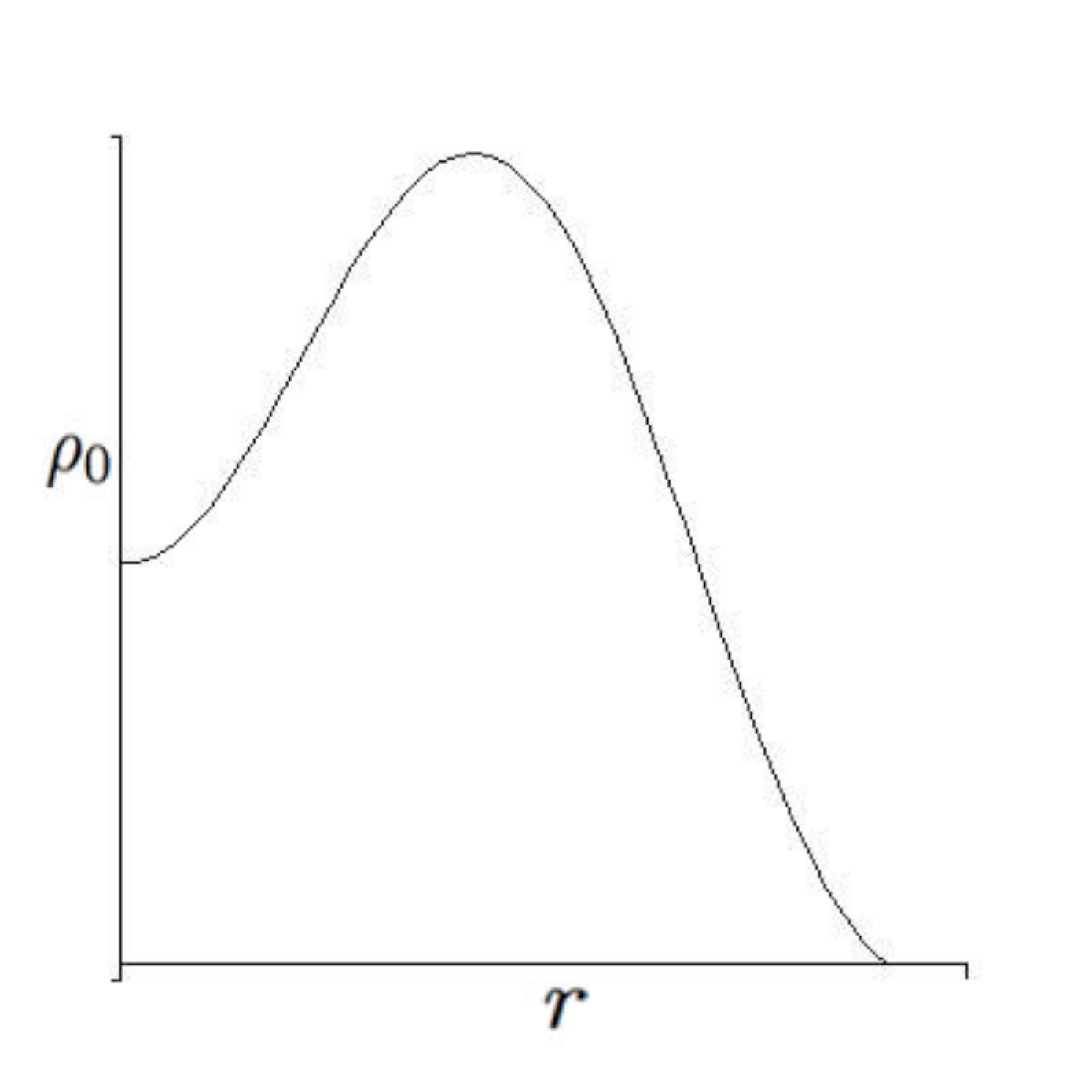}
		\caption{\label{rhoplot} Initial density profile}
	\end{center}
\end{figure}
While the density profile given (Fig. \ref{rhoplot}) is an unrealistic starting point for gravitational collapse, we note that the energy density in realistic collapse scenarios may not necessarily be monotonically decreasing, and may therefore have corresponding regions where the tidal force takes on the opposite sign.  For example, an accretion disc around a black hole will have a distribution not dissimilar to the present example.    

\begin{figure}
	\begin{center}
		\includegraphics[height=0.45\textwidth,width=0.45\textwidth]{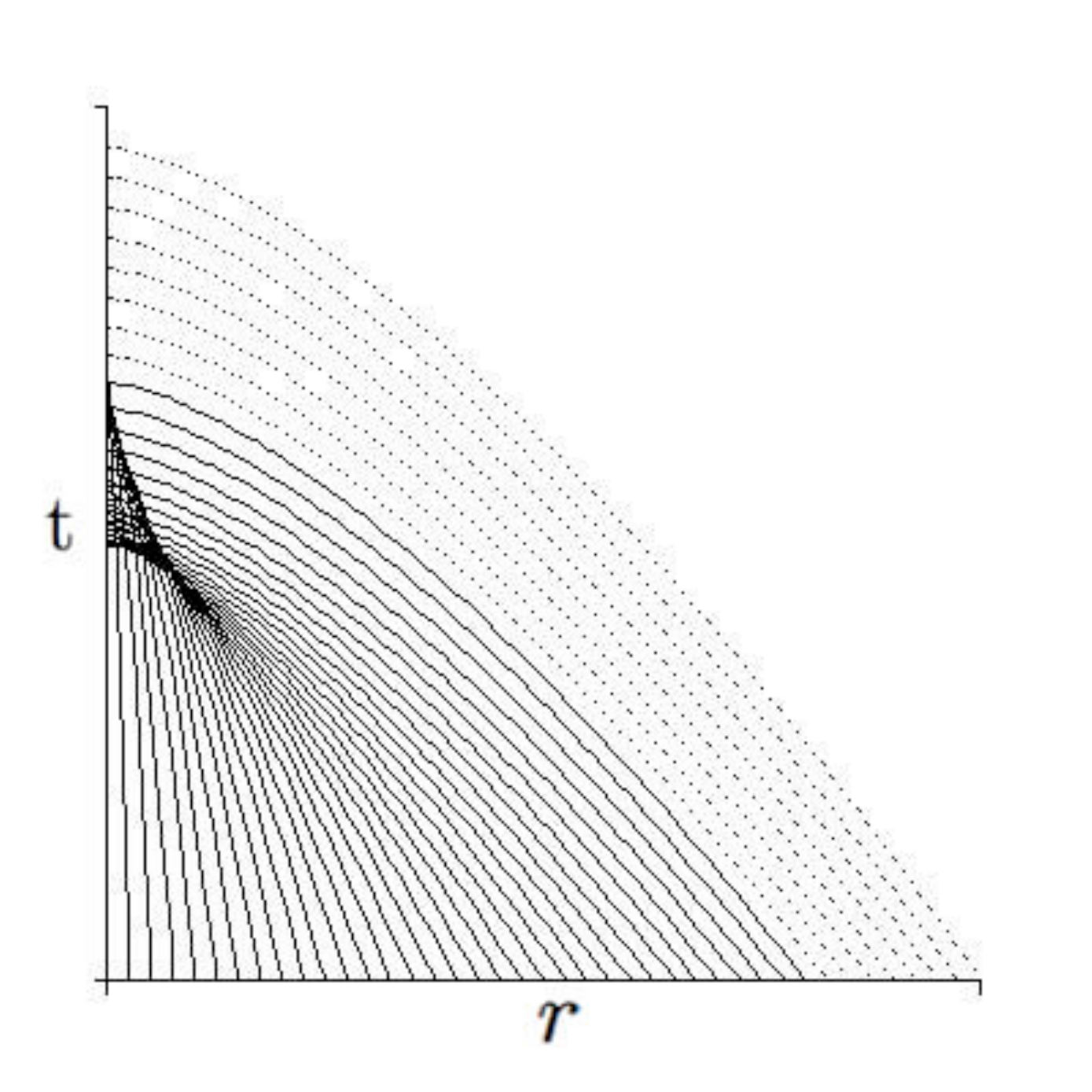}
		\caption{\label{roo}  Characteristic curves of constant mass}
	\end{center}
\end{figure}

\begin{figure}
	\begin{minipage}{1\textwidth}
		\begin{center}
		\includegraphics[height=0.3\textwidth,width=0.3\textwidth]{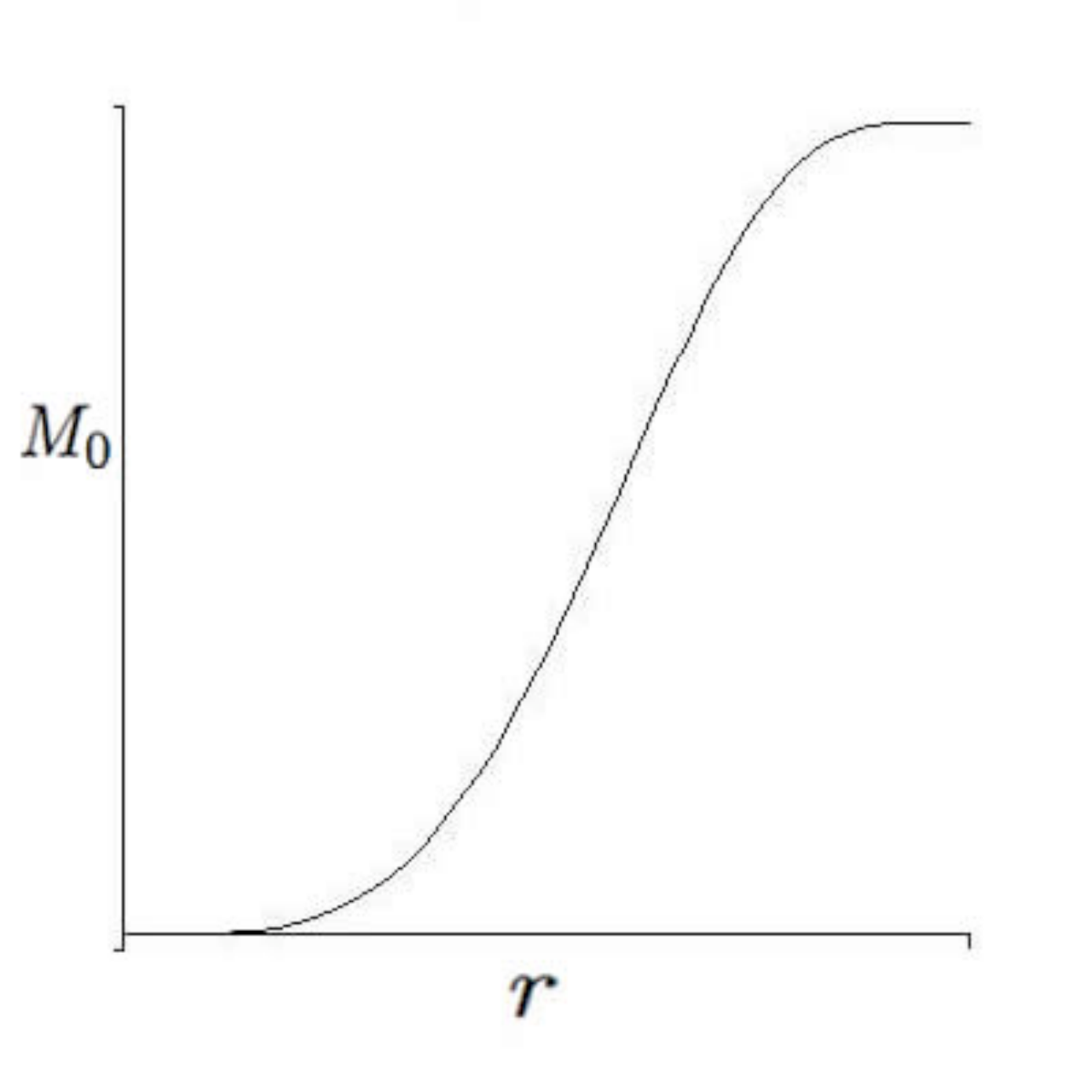}
		\includegraphics[height=0.3\textwidth,width=0.3\textwidth]{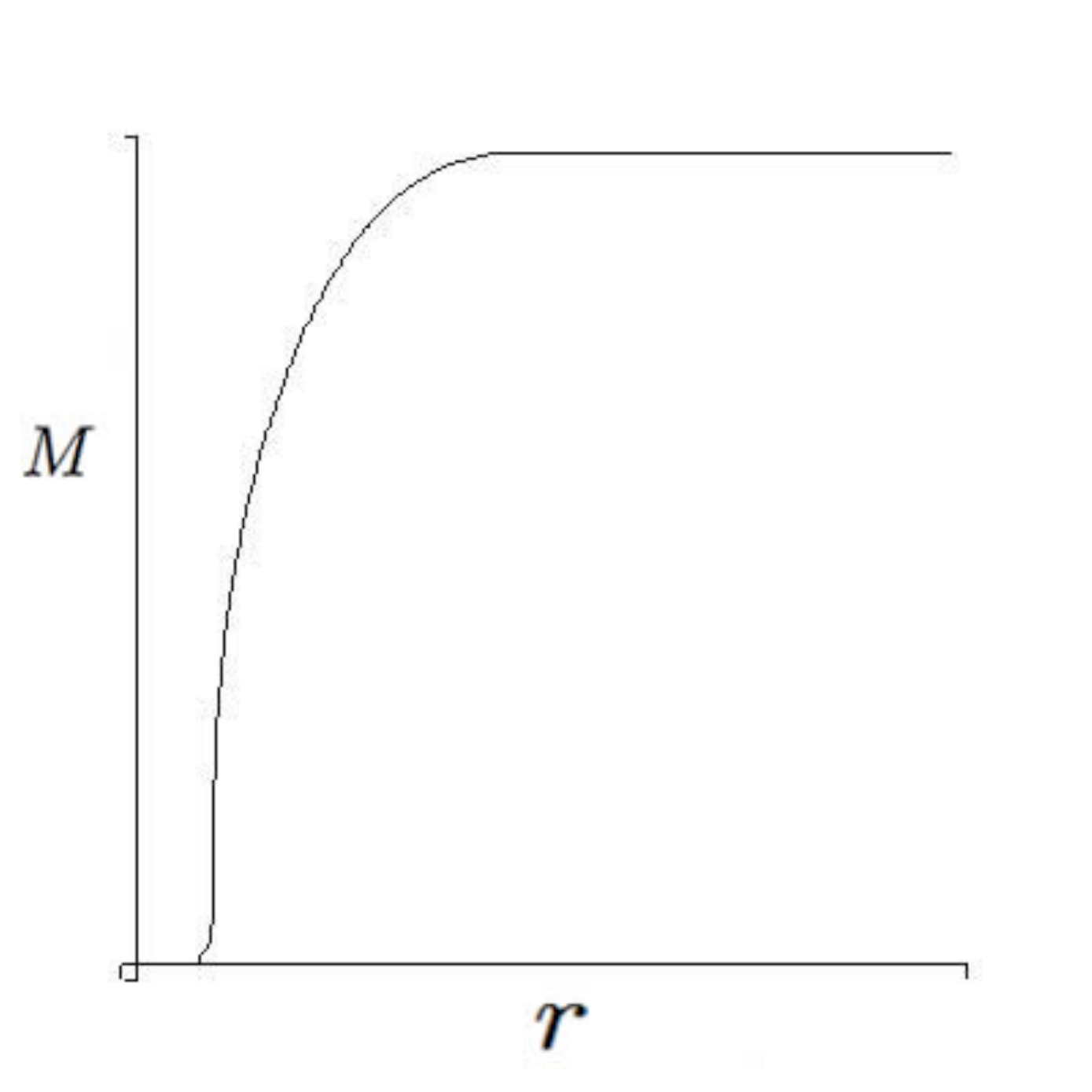}
		\includegraphics[height=0.3\textwidth,width=0.3\textwidth]{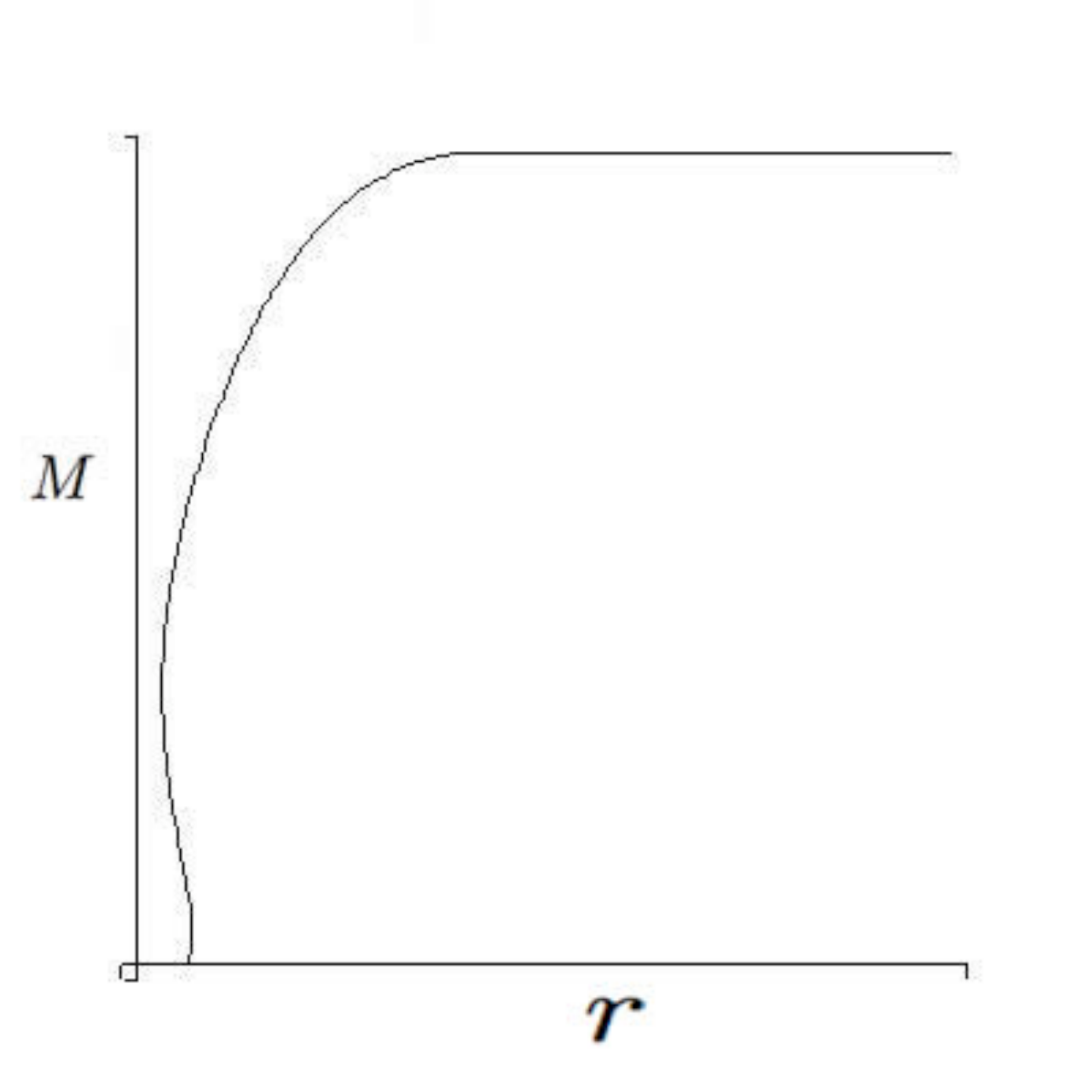}
		\caption{\label{muhyhu} Initial Mass, Mass profile at shock formation, Mass profile after shock formation}
	\end{center}
	\end{minipage}
\end{figure}

While an analytic solution utilizing the above initial conditions can be found, it is extremely long, implicit and highly non-linear.  However, the classical solution can be plotted even past the initial formation of the shock.  We do this only to display the nature of the shell crossing singularity being a multi-valued function in the mass.   Figure \ref{roo} displays the characteristic curves of constant mass emanating from the initial distribution.  The dashed curves are those coming from the exterior, vacuum region of the spacetime.  The shock wave is represented here by a crossing of the characteristic curves.     
Figure \ref{muhyhu} displays three plots of the Mass function against the radius of the classical solution at three different times.  

\section{Conclusion}

A review of the literature associated with LTB collapse shows that major phenomenological discussion is focused on SCS (e.g., \cite{christodoulou84,newman86,szekeres99,nolan02}).  However, much of this work is done in terms of arbitrary functions which are difficult to interpret.  

The work herein derived LTB regions and Schwarzschild regions in GPG coordinates via an initial value formulation.  The advantage of this approach is four-fold:  
\begin{enumerate} 
	\item It enabled the metric to be written in terms of a single line element, thus avoiding complicated matching schemes at the interface.  This differs from the standard approach in the literature (e.g., \cite{misner73}) where two spacetimes are matched across a boundary by a coordinate transformation.  Difficulties often arise in showing that this coordinate transformation is valid everywhere along the boundary as the Jacobian of transformation is non-trivial to analyse.  By writing the line element in a single coordinate patch, no transformation is required and this difficulty is avoided.  It has already been shown that this method is extendable beyond the dust case to include both perfect fluids \cite{lasky06b} and also a completely general fluid \cite{lasky07}.  However, the equations derived in those cases are extremely difficult to handle due to more terms in the fluid implying more terms are also required to describe the geometry (see \cite{lasky07} for a discussion of this point).  The dust cases analysed in this article provides relatively straightforward analytic solutions, and hence the structure of the equations is better understood.    
	\item Furthermore, as analytic solutions are found from the initial conditions, this approach is excellent as a test-bed for numerical schemes.  In particular, the evolution of the boundary of the two regions of the spacetime and of the SCS singularity can be tested as exact analytic forms of both of these are known.   
	\item The metric coefficients are all expressed in terms of the matter fields.  Furthermore, the dynamics of the spacetime is completely determined by two differential equations governing the mass and energy functions.  The problem is then solved using {\it physically reasonable} initial and boundary conditions on the energy density.  The evolution of this mass function can result in multi-valued solutions for particular choices of initial conditions, which is exactly equivalent to SCS.  The advantage of this scheme is that it explicitly identifies the SCS to be equivalent to shock waves.  This enables one to analyse the dynamical extensions of SCS beyond their initial point of formation.  In particular, we showed that a SCS that forms as a globally naked singularity must become a locally naked singularity, and eventually fall to a Schwarzschild singularity at $r=0$, at which point it will cease to effect the system.
	\item The spacetime is easier to visualize due to the physical intuition being similar to familiar fluid problems.  Rather than dealing with abstract four-dimensional coordinates, we deal with the propagation of hypersurfaces, exactly as we imagine the spacetime in which we live.  Mathematically, this implies all functions appearing in the solution are not necessarily arbitrary, but are more appropriately based on physical intuition. 
\end{enumerate}

\section*{Acknowledgements}
The authors wish to thank Brien Nolan for useful discussions.  All calculations were checked using the computer algebra program Maple.    

\renewcommand{\theequation}{A\arabic{equation}}
\setcounter{equation}{0}  
\renewcommand{\thesection}{\arabic{section}}
\setcounter{section}{0}
\section*{Appendix}\label{App}  

For clarity, we express the extrinsic curvatures $K_{ij}$, and three-Riemann curvatures ${^3R}_{ij}$, in terms of the metric coefficients.
The Lie derivative operator with respect to the normal $\mathcal{L}_n$, is expressed when operating on a scalar in equation ($\ref{Liedef}$), and we have already let $\alpha=1$ (see section \ref{ADM}).

{\singlespacing
\begin{align}
{K_i}^j=\left(\begin{array}{ccc}
\frac{\partial \beta}{\partial r}-\frac{1}{\mathcal{U}}\mathcal{L}_n\mathcal{U} & 0 & 0\\
0 & \frac{\beta}{r}& 0\\
0 & 0 & \frac{\beta}{r}
\end{array}\right),
\end{align}
\begin{align}
K=\frac{1}{r^2}\frac{\partial}{\partial r}\left(\beta r^2\right)-\frac{1}{\mathcal{U}}\mathcal{L}_n\mathcal{U},
\end{align}
\begin{align}
{A_i}^j=\frac{1}{3}\left[r\frac{\partial}{\partial r}\left(\frac{\beta}{r}\right)-\frac{1}{\mathcal{U}}\mathcal{L}_n\mathcal{U}\right]\left(\begin{array}{ccc}
2 & 0 & 0\\
0 & -1 & 0\\
0 & 0 & -1
\end{array}\right),
\end{align}
\begin{align}
{{^3R}_i}^j=\frac{1}{r\mathcal{U}^3}\left(\begin{array}{ccc}
2\frac{\partial\mathcal{U}}{\partial r} & 0 & 0\\
0 & \frac{\partial\mathcal{U}}{\partial r}+\frac{1}{r}\left(\mathcal{U}^3-\mathcal{U}\right) & 0\\
0 & 0 & \frac{\partial\mathcal{U}}{\partial r}+\frac{1}{r}\left(\mathcal{U}^3-\mathcal{U}\right)
\end{array}\right),
\end{align}
\begin{align}
{^3R}=\frac{2}{r\mathcal{U}^3}\left[2\frac{\partial\mathcal{U}}{\partial r}+\frac{1}{r}\left(\mathcal{U}^3-\mathcal{U}\right)\right].
\end{align}
}

\bibliographystyle{anziam}
\bibliography{Dust}

\end{document}